\documentclass{article}

\usepackage{PRIMEarxiv}

\usepackage[utf8]{inputenc} 
\usepackage[T1]{fontenc}    
\usepackage{hyperref}       
\usepackage{url}            
\usepackage{booktabs}       
\usepackage{amsfonts}       
\usepackage{nicefrac}       
\usepackage{microtype}      
\usepackage{lipsum}
\usepackage{fancyhdr}       
\usepackage{graphicx}       
\usepackage{orcidlink}
\graphicspath{{media/}}     

\title{Deformation monitoring with Sentinel-1 Wave mode data
\thanks{\textit{\underline{Email}}: 
\texttt{piyush@descarteslabs.com}} 
}

\author{Piyush S. Agram\, \orcidlink{0000-0003-0711-0264},
       Matthew T. Calef\, \orcidlink{0000-0003-4701-7224},
       Kelly M. Olsen\, \orcidlink{0000-0002-2709-9237},
       Kimberly Carlson\, \orcidlink{0000-0001-9154-1248},
       Scott Arko \\
       Descartes Labs Inc.\\
       Santa Fe, NM, USA}

\begin{document}
\maketitle

\begin{abstract}
We describe the salient characteristics of Sentinel-1 wave (WV) mode vignettes. We describe our approach for working with WV mode data that enables vignette-based data access and processing, thereby eliminating the Sentinel-1 Single Look Complex (SLC) data packaging and current archive metadata conventions as a bottleneck to large scale processing. We discuss the spatial and temporal coverage of Sentinel-1 WV mode data and show that a large volume of data has been acquired over land masses in this mode, thus allowing us to use it for land monitoring applications as well as ocean applications. For targeted infrastructure monitoring studies, we are able to generate coregistered, geocoded stacks of WV mode SLCs for any area of interest (AOI) with sufficient wave mode coverage, in a few minutes. We demonstrate the applicability of using WV mode data for deformation monitoring applications. Finally, we discuss the benefits and limitations of working with Sentinel-1 WV mode data.
\end{abstract}


\section{Introduction}
\label{sec:intro}

The Sentinel-1 synthetic aperture radar (SAR) constellation~\cite{Torres:2012}, part of Europe’s Copernicus Earth Observation programme, has been operational since October 2014 and has significantly boosted the scientific community's  ability to monitor our planet's surface at a global scale. Data acquired by this constellation have been freely available for end users, enabling local and global scale application users to reliably access and analyze SAR imagery over their areas of interest (AOIs). A number of wide-area land monitoring applications have been built on top of Sentinel-1's data archive, e.g.,~\cite{Ferretti:2021}. 

As described in~\cite{Torres:2012}, the operation of the Sentinel-1 satellites follows a pre-programmed conflict-free observation plan where the imaging modes, i.e. Interferometric Wide (IW) swath mode and Extended Wide (EW) swath mode,  can be operated for a maximum of 25 minutes per orbit. For the remaining time the instrument operates over the open ocean in the Wave (WV) mode providing sampled images of 20 × 20 km area, known as {\it vignettes}, every 100 km along the orbit at a low data rate mode (see Figure~\ref{fig:safe}). While the use of IW mode Sentinel-1 imagery is well established for various applications using both radar backscatter and interferometry, e.g.,~\cite{Ferretti:2021,Agram:2022}, the characteristics of the WV mode are less well known. In this manuscript, we describe the salient features of this mode and its applications in detail.

This manuscript is organized as follows. Section~\ref{sec:wave} provides an overview of Sentinel-1's WV mode, including a detailed look at its spatio-temporal coverage characteristics. We present a short overview of possible applications using Sentinel-1 WV mode data in Section~\ref{sec:examples}. In Section~\ref{sec:insar}, we delve deeper into the use of WV mode data for interferometry and deformation monitoring applications. Finally, in Section~\ref{sec:conclusion}, we discuss the benefits and limitations of using Sentinel-1 WV mode data for land monitoring applications.

\section{Sentinel-1 WV mode}
\label{sec:wave}

The WV imaging mode~\cite{Lehner:2000,Hasselmann:2013} was specifically designed to measure ocean wave spectra and ocean surface winds at a global scale. The Sentinel-1 mission acquires WV mode vignettes in a  "leapfrog" pattern~\cite{S1userguide:2023}, i.e, data is acquired in a single polarization (VV or HH) over 20 km by 20 km vignettes every 100 km along the orbit, acquired alternately on two different beams or incidence angles. Vignettes on the same beam are separated by 200 km. Swaths alternate beams between near range (WV1) and far range (WV2). The detailed imaging characteristics of the WV mode beams are enumerated in Table~\ref{tab:modeparams}. Note that WV mode has a higher spatial resolution (4 to 5 meters) compared to the operational IW and EW modes. 
\begin{table}[!h]
 \caption{Sentinel-1 WV mode swath characteristics}
  \centering
  \begin{tabular}{lll}
    \toprule
     & WV1     & WV2 \\
    \midrule
    Wavelength & $5.5466$ cm & $5.5466$ cm \\
    Antenna length & $12.3$ m & $12.3$ m \\
    Bandwidth & $74.5$ MHz  & $48.2$ MHz     \\
    Range sampling rate & $100.1$ MHz & $54.6$ MHz \\
    Pulse repetition frequency & $1.65$ KHz & $1.65$ KHz \\ 
    Incidence angle at minimum altitude    & $23.47^{\circ}$ - $25.03^{\circ}$ & $36.67^{\circ}$ - $37.92^{\circ}$      \\
    Look angle at minimum altitude     & $21.03^{\circ}$ - $22.40^{\circ}$       & $32.56^{\circ}$ - $33.62^{\circ}$  \\
    Incidence angle at maximum altitude    & $21.68^{\circ}$ - $23.22^{\circ}$ & $34.88^{\circ}$ - $36.13^{\circ}$  \\
    Look angle at maximum altitude     & $19.43^{\circ}$ - $20.79^{\circ}$       & $30.96^{\circ}$ - $32.02^{\circ}$  \\
    \bottomrule
  \end{tabular}
  \label{tab:modeparams}
\end{table}

It is also important to understand that WV mode is a lower "data" rate imaging mode~\cite{Torres:2012} and hence has a lower signal-to-noise (SNR) ratio than the operational imaging modes - IW and EW. The Sentinel-1 Mission Performance Cluster (\url{https://sar-mpc.eu/}) consistently tracks the performance of this imaging mode and reports that the Normalized Radar Cross Section (NRCS) bias is near zero dB and the standard deviation is 1.6 dB for WV1 and 1.8 dB for WV2 beams~\cite{Hajduch:2022}. It should also be noted that the WV2 antenna pattern originally used the stripmap S4 beam coefficients due to a limit on the number of possible patterns in the on-board memory resulting in worse performance in terms of Noise Equivalent Sigma Zero (NESZ). An optimised WV2 antenna pattern configuration was implemented on the 22nd June, 2021 for Sentinel-1A and on the 24th June, 2021 for Sentinel-1B allowing a NESZ improvement of about 5 dB~\cite{Hajduch:2022}.

\subsection{Data packaging}
\label{sec:zipfile}

\begin{figure}[!h]
  \centering
  \includegraphics[height=6cm]{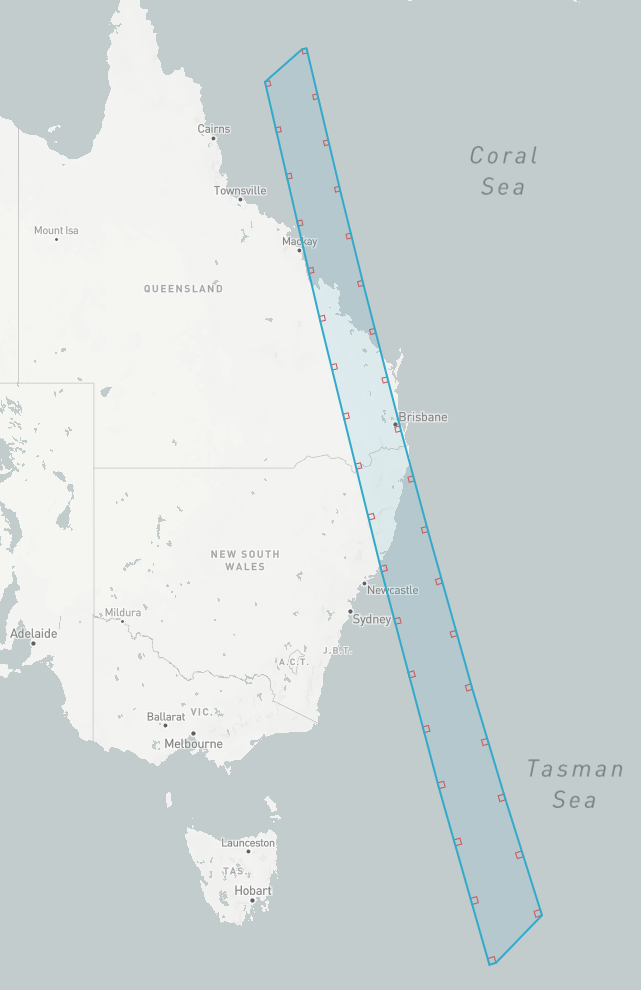}
  \caption{Geometry associated with a single WV mode SAFE granule (Blue) in the Copernicus Data Space Ecosystem (CDSE) archive and the actual bounding boxes of individual vignettes (Red) included in the granule. Note that the SAFE granule contains no usable data over the most of the geometry associated with the granules in the distribution archive.}
  \label{fig:safe}
\end{figure}

Sentinel-1 WV mode SLC data are distributed in the Standard Archive Format for Europe (SAFE) format, with each vignette included as a stand-alone TIFF file within the SAFE granule. All vignettes acquired in a single datatake are packaged together in a single SAFE granule, resulting in a large variation in the size of these granules. A single WV mode SLC SAFE granule can contain from anywhere between 15 to 160 vignettes. The SAFE granules are also tagged with the geometry of the bounding box of all contained vignettes as shown in Figure~\ref{fig:safe}.

Following~\cite{Agram:2022}, we have labelled and indexed every individual vignette in WV mode SLCs archived at the Alaska Satellite Facility (ASF), and built data access mechanisms that avoid having to download entire SAFE archives. This allows us to scale our processing pipelines with WV mode data in a manner similar to our burst-based IW mode pipelines. Sentinel-1 WV mode acquisitions are not synchronized~\cite{Torres:2012}, and hence do not follow a consistent underlying tiling system like the IW mode bursts~\cite{Agram:2022, Burst:2022}. Nevertheless, knowing the location of each individual vignette allows us to treat this dataset like those provided by tasked commercial providers and access them on-demand at scale.
 
\subsection{Acquisition history}

\begin{figure}[!h]
  \centering
  \includegraphics[height=6cm]{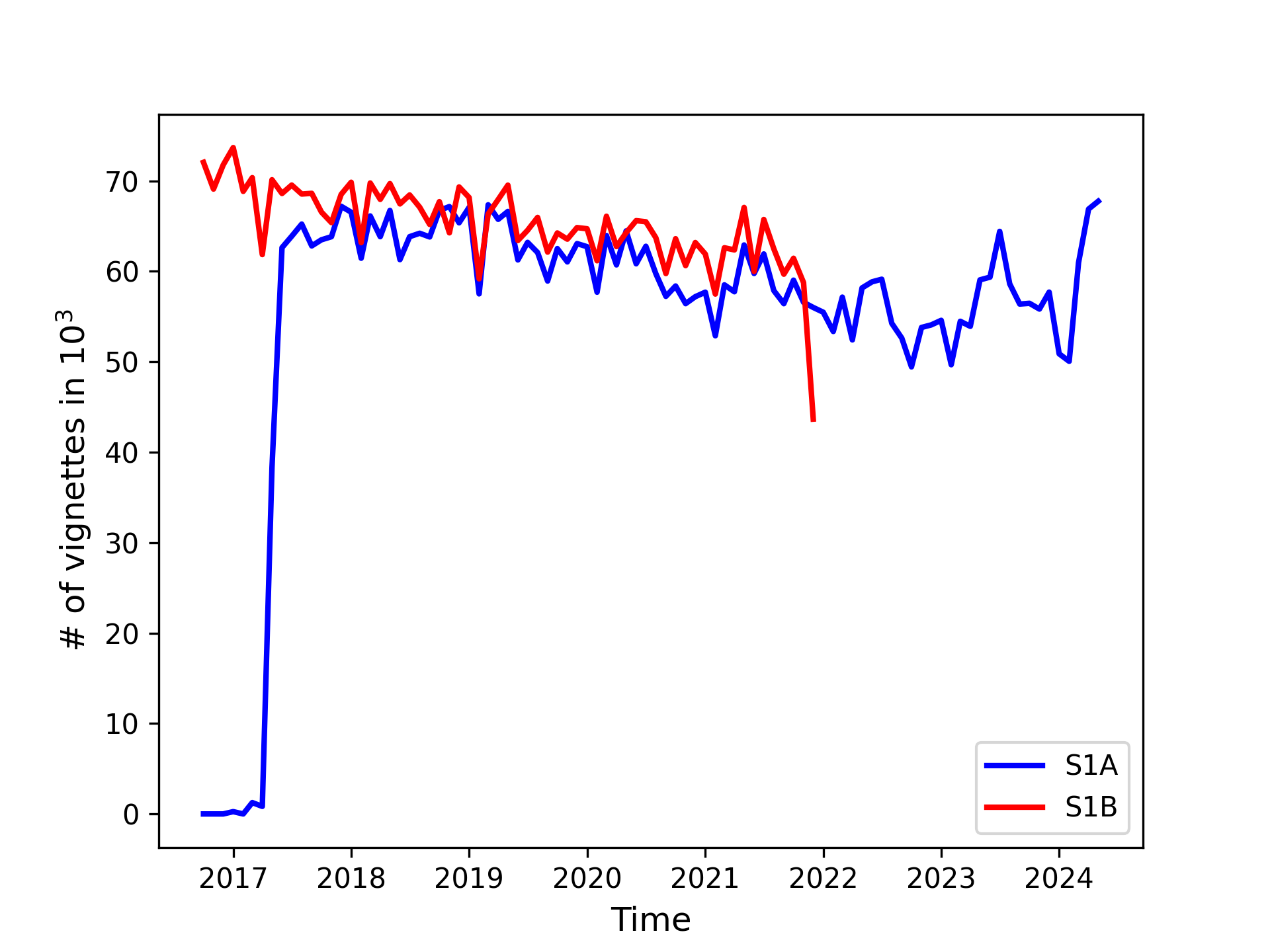}
  \caption{Number of vignettes per month over the history of the Sentinel-1 mission through June 1, 2024. These are accessible via the Alaska Satellite Facility (ASF) archive.}
  \label{fig:count}
\end{figure}

Figure~\ref{fig:count} shows the count of vignettes acquired by each satellite per month over the history of the Sentinel-1 mission, that is available via the ASF archive. Over nine million individual vignettes have been acquired so far at a rate of about sixty thousand vignettes per satellite per month. From Figure~\ref{fig:count}, it appears that the number of WV mode acquisitions using Sentinel-1A declined after the Sentinel-1B outage in December 2021, possibly due to increased demand for coverage in IW mode. We would expect the number of WV mode acquisitions to recover to nominal levels again after the launch of Sentinel-1C.

\subsection{Coverage of land masses}
\label{sec:landcoverage}

\begin{table}[!h]
 \caption{Approximate vignette count per continent over the history of the Sentinel-1 mission untill June 1, 2024.}
  \centering
  \begin{tabular}{lc}
    \toprule
     Continent     & Approximate Count \\
    \midrule
    Africa & $65000$\\
    Australia & $34000$  \\
    North America & $40000$ \\
    South America & $45500$\\
    \midrule
    Total & $183500$\\
    \bottomrule
  \end{tabular}
  \label{tab:continent}
\end{table}

\begin{figure}[!h]
  \centering
  \includegraphics[width=0.45\linewidth]{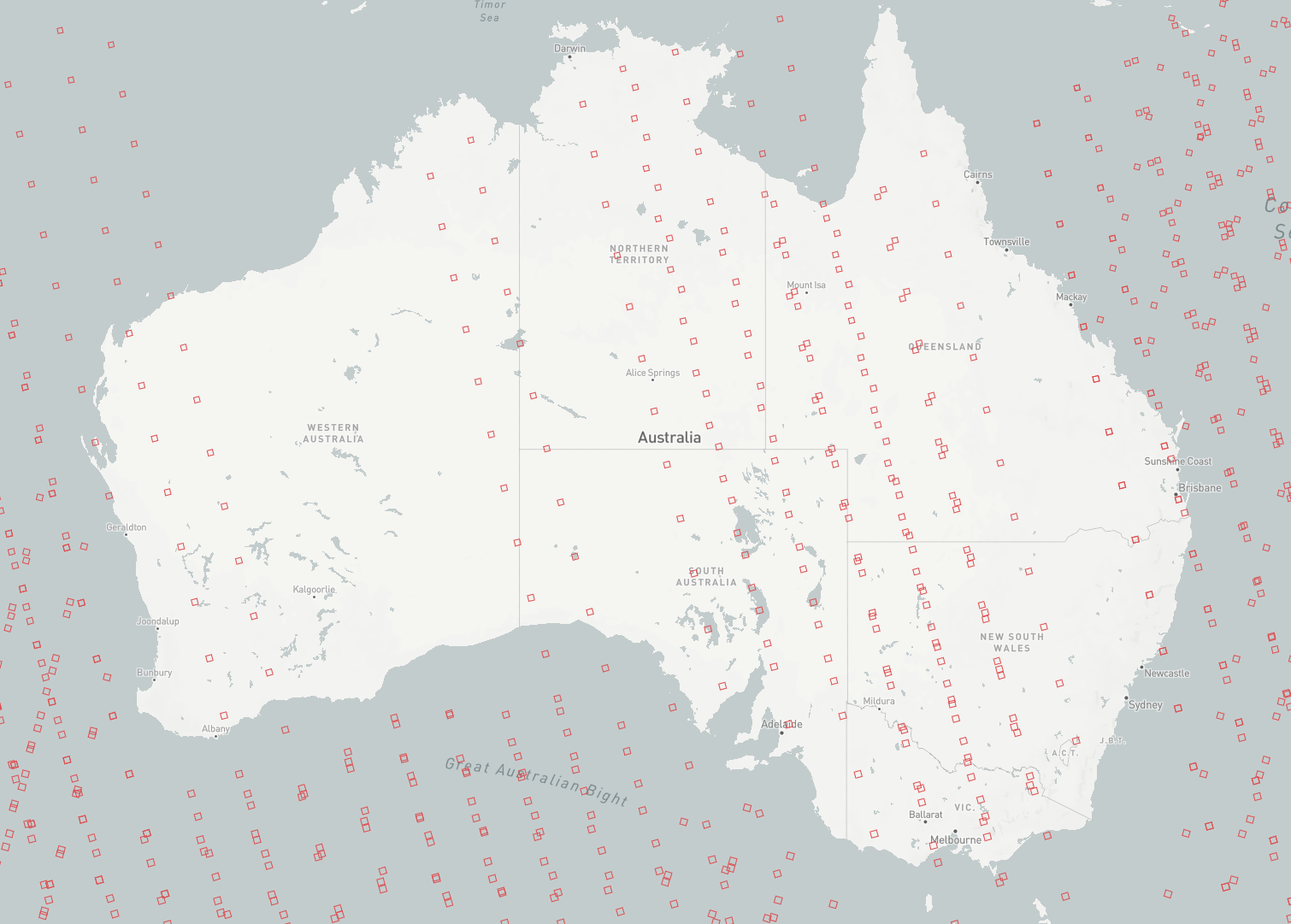}
  \includegraphics[width=0.45\linewidth]{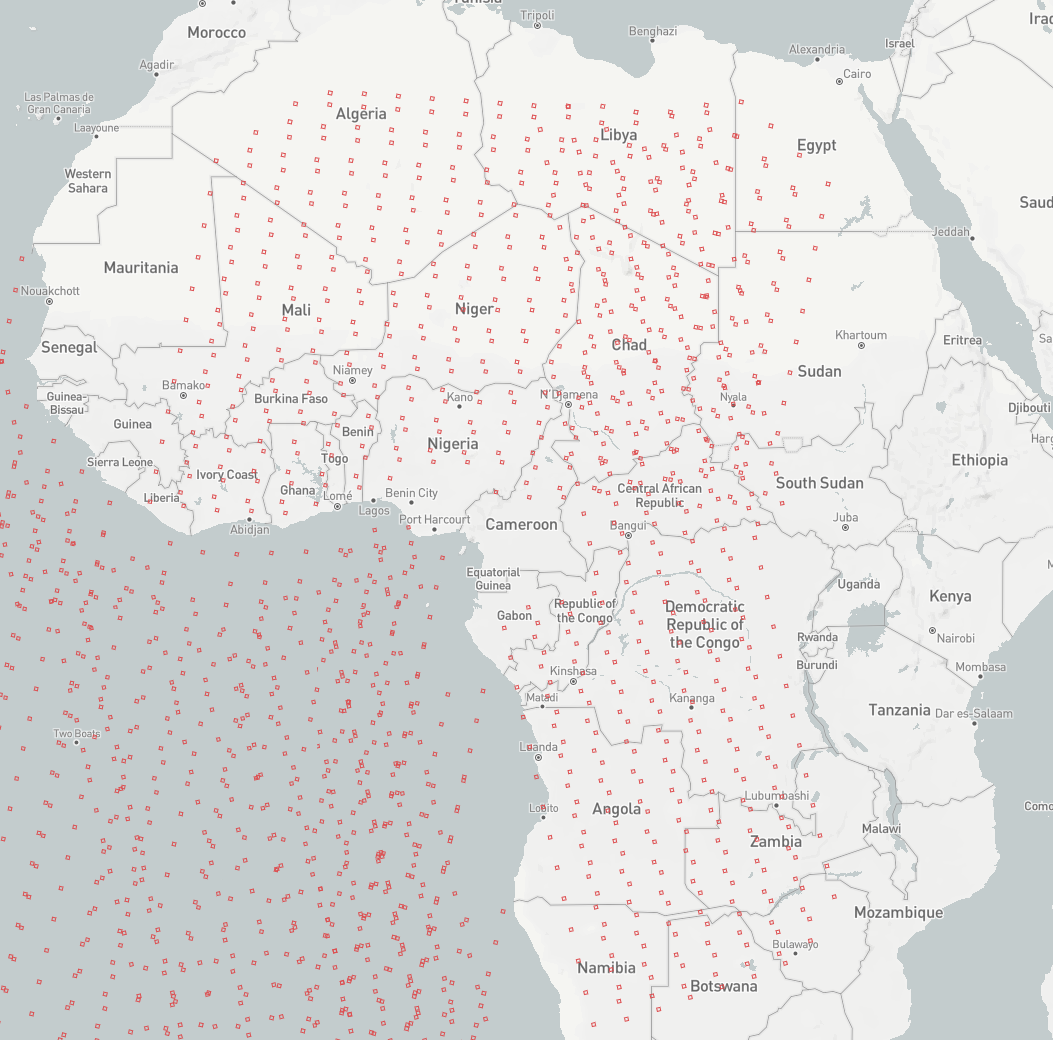}
  \caption{Footprints of vignettes acquired between May 1, 2020 to May 13, 2020  over Australia (Left) and Africa (Right), that are accessible via the Alaska Satellite Facility (ASF) archive.}
  \label{fig:continents}
\end{figure}

The WV mode acquisitions are not synchronized~\cite{Torres:2012}. However, because the Sentinel-1 mission follows a conflict-free observation scenario that is not modified often, we fortuitously end up with deep stacks of repeating vignettes. When both Sentinel-1A and Sentinel-1B satellites were in operation, more data was acquired over land masses. We present examples of WV mode coverage over a 12-day cycle in May 2020 over Australia and Africa in Figure~\ref{fig:continents}. Table~\ref{tab:continent} shows the approximate number of vignettes over different continental land masses in the archive so far. WV mode coverage over mainland Asia and Europe is almost zero and negligible over Antarctica. Note that a vast majority of these vignettes over land masses were acquired repeatedly due to the mission's stable observation scenario. Since the Sentinel-1B outage in December 2021, WV mode coverage over land masses is largely restricted to Eastern Australia.

\section{Monitoring applications using WV mode}
\label{sec:examples}

\begin{figure}[!h]
  \centering
\includegraphics[width=0.8\linewidth]{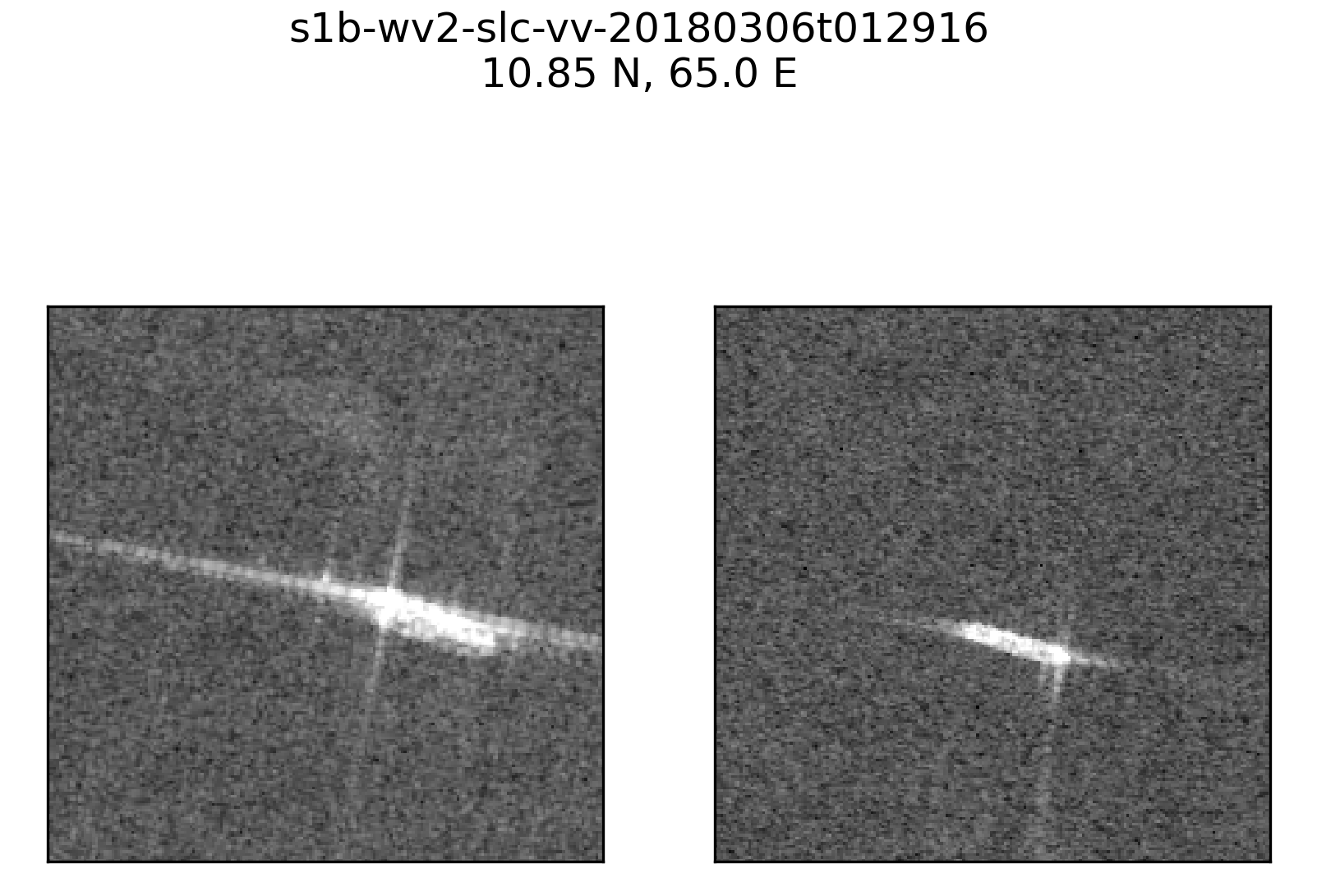}
  \caption{Two 1000m x 1000m chips from a single vignette showing bright targets in the Arabian sea in March 2018 in 5 meter WV mode backscatter imagery. This region is not imaged nominally by optical satellites or other SAR satellites.}
  \label{fig:arabiansea}
\end{figure}

In this section, we present an overview of possible applications with Sentinel-1 WV mode data. The WV mode was designed and implemented to study ocean winds and waves~\cite{Lehner:2000,Hasselmann:2013}. The associated Sentinel-1 Level 2 Ocean (OCN) products~\cite{S1userguide:2023}, their validation and applications are addressed in great detail in the literature of oceanography and ocean remote sensing communities, e.g., ~\cite{Ardhuin:2015, Stopa:2017, Li:2019,Wang:2019} and we refer interested readers to \url{https://oceanwavesremotesensing.ifremer.fr/} for these topics. Instead, we treat Sentinel-1 WV mode SLCs like SAR imagery acquired by any other sensor operating in stripmap mode and focus on its applicability to traditional SAR applications. We provide some simple examples and do not delve into implementation details, as deformation monitoring (Section~\ref{sec:insar}) is the primary focus of this manuscript.

\subsection{Backscatter-based ocean applications}
\label{sec:oceanapp}

\begin{figure}[!h]
  \centering
  \includegraphics[width=0.9\linewidth]{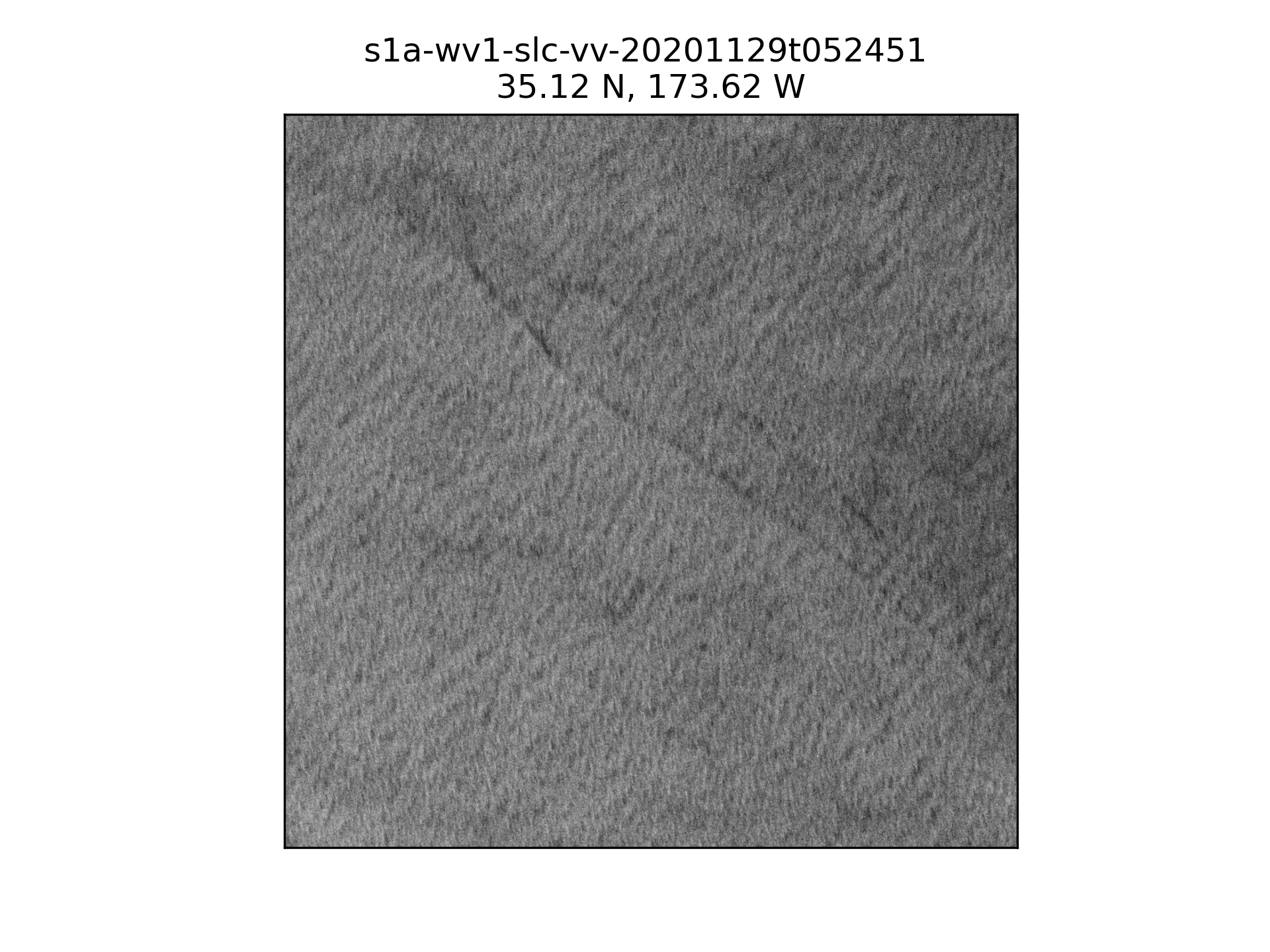}
  \caption{Possible oil related artifacts in a 6000 meter-by-6000 meter parch of the Pacific ocean in November 2020. This region is not imaged nominally by optical satellites or other SAR satellites regularly.}
  \label{fig:maritime}
\end{figure}

Sentinel-1 WV mode data is, as far as we know, the only public source of imagery with approximately five meter resolution that is consistently acquired over open oceans where traditional remote sensing missions like Landsat and Sentinel-2 usually don't acquire imagery. Figures~\ref{fig:arabiansea} and~\ref{fig:maritime} includes examples of direct and indirect observations of maritime activity over remote regions that are far away from the coastline to be regularly monitored using other publicly available datasets. In Figure~\ref{fig:arabiansea}, we include an image in the Arabian sea over a relatively busy shipping lane from March 2018. In Figure~\ref{fig:maritime}, we include a vignette with no obvious bright targets but showing possible oil-related streaks from maritime activity on a busy shipping lane in the Pacific ocean from November 2020. 

\subsection{Backscatter-based land applications}
\label{sec:sigma}

\begin{figure}[!h]
  \centering
  \includegraphics[width=0.5\linewidth]{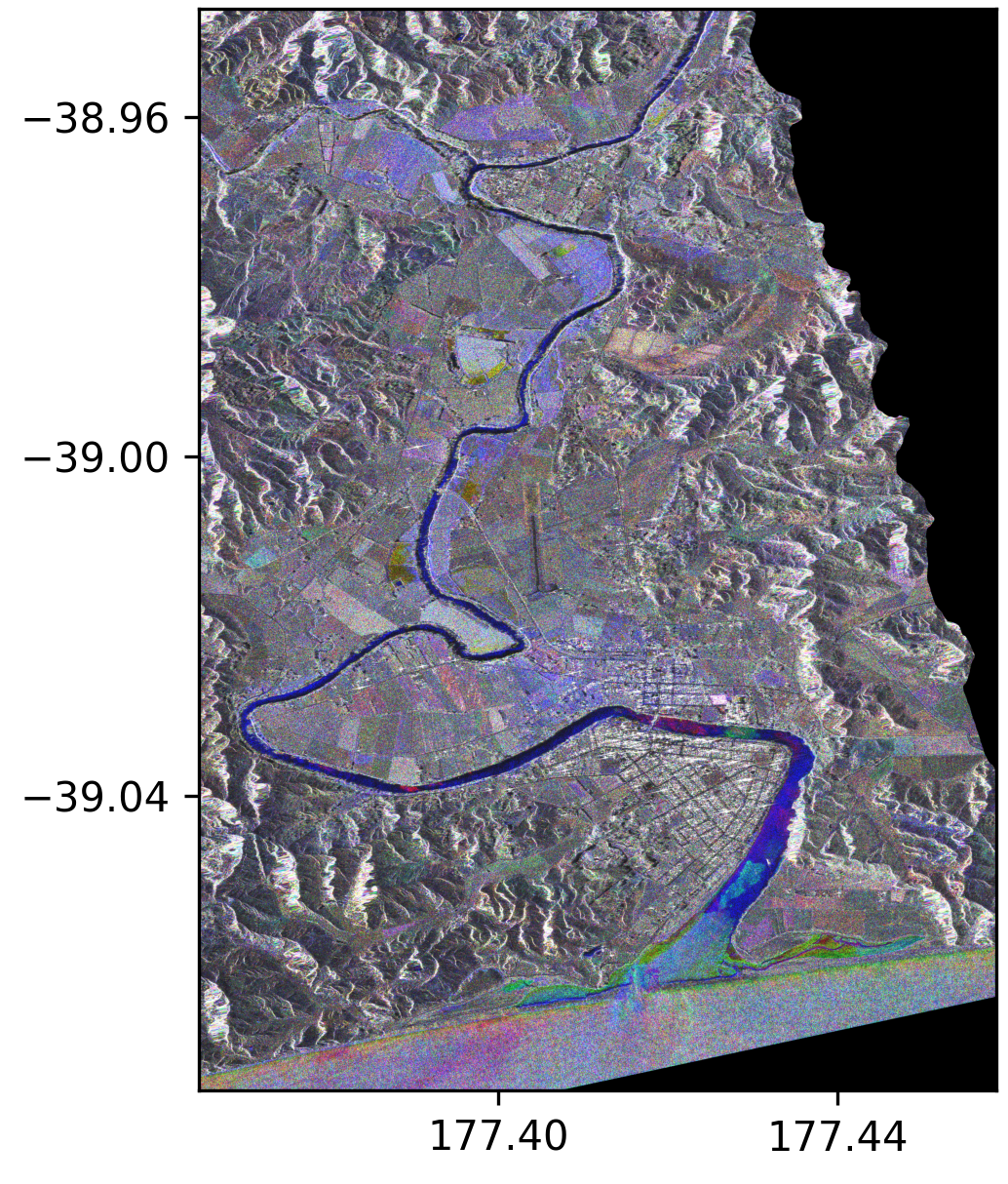}
  \caption{Simple RGB composite of three WV mode vignettes from Track 110 acquired on Jan 23, Feb 4 and Feb 16 2023 over Wairoa, New Zealand. This area was impacted by Cyclone Gabrielle in Feb 2023.}
  \label{fig:wairoa}
\end{figure}

We process WV mode vignettes on to a 5 meter grid using the same pipelines that we have developed for IW bursts~\cite{Agram:2022}. We do observe along-track offsets in geolocation for WV mode and we address this in Section~\ref{sec:coreg}. In general, we are able to use WV mode backscatter data in the same manner as IW mode backscatter data. We only present one example here. Figure~\ref{fig:wairoa} shows a simple RGB composite of three vignettes from Track 110 acquired on Jan 23\slash Feb 4\slash Feb 16 2023 over Wairoa, New Zealand. This area was impacted by Cyclone Gabrielle on this time frame and the affected areas are discernible. Though this area was also imaged in IW modes on Track 8 and Track 175, these acquisitions are only 24 hours apart resulting in 11-12 day time gap between revisits. WV mode data complements the IW acquisitions and improves the temporal sampling for monitoring applications and disaster response.

\subsection{Proxy for commercial smallsat data}
\label{sec:proxy}

Sentinel-1 WV mode data appears to have SNR and geolocation accuracy characteristics that are similar to current smallsat SAR satellites like Iceye, Capella and Umbra. While only a few hundreds of scenes from these smallsat sensors are available in the public domain to develop pipelines and models, the nine million WV mode vignettes acquired so far can help fill the gap for many applications. While not perfect, the higher resolution of WV mode data is a better proxy for smallsat data than the lower resolution IW mode.

\section{Deformation monitoring using WV mode}
\label{sec:insar}

In this section, we delve deeper into geometric properties of Sentinel-1 WV mode and our approach to generating coregistered stacks from this data. We also present an example of deformation monitoring using WV mode data over a mining area. We observe that while WV mode data has lower SNR than the operational imaging modes from Sentinel-1, it still benefits from the mission's narrow orbit tube~\cite{Torres:2012}. Specifically, if vignettes from the same track overlap on repeat passes they are almost guaranteed to support InSAR analysis.

\subsection{Geometric accuracy}
\label{sec:geomcalib}

Relative geometric accuracy for the Sentinel-1 IW mode, which is the primary driver for automating interferometric and change detection analysis, is on the order of 15-20cm~\cite{Mirandageo:2017}. This is an order of magnitude less than the resolution of the IW mode SAR imagery, allowing us to geocode SLC data directly and significantly simplify SAR-InSAR processing workflows~\cite{Agram:2022}. While several studies have focused on the geometric accuracy of IW and stripmap modes of Sentinel-1, we did not find any information related to the geometric accuracy of WV mode in literature.

\begin{figure}[!h]
  \centering
  \includegraphics[width=0.45\linewidth]{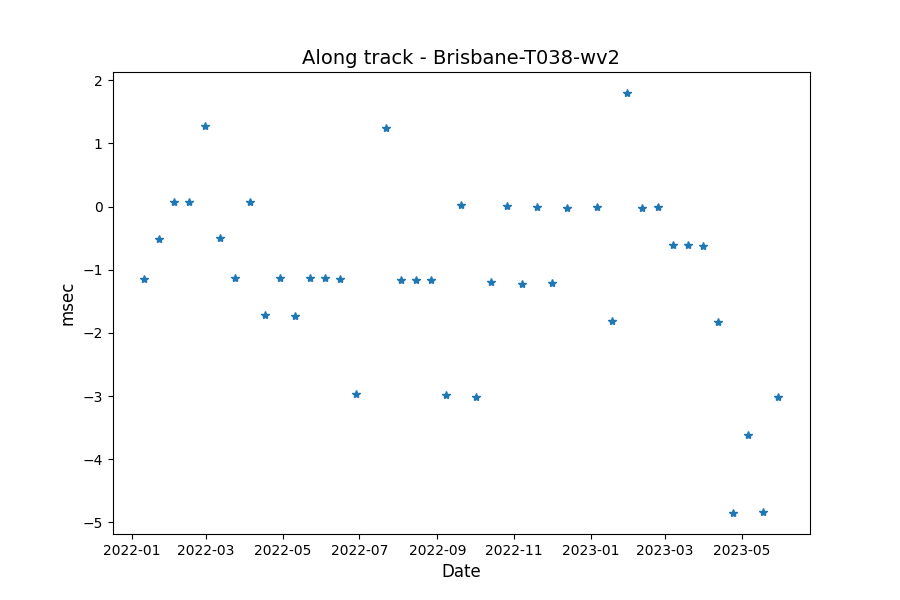}
  \includegraphics[width=0.45\linewidth]{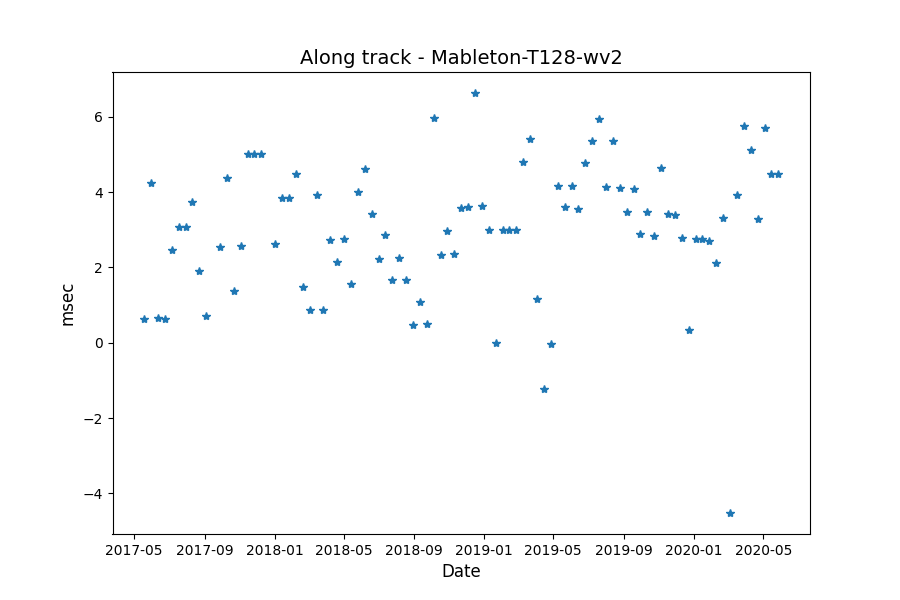}
  \caption{Relative along track time shifts in milliseconds, with respected to an arbitrarily chosen reference scene in the stack, estimated using image cross-correlation for two different stacks over (Left) Brisbane, Queensland, Australia and (Right) Mableton, Georgia, USA.}
  \label{fig:offsets}
\end{figure}

Since, we do not have access to corner reflectors located within WV mode vignette footprints for characterizing geometric accuracy, we apply techniques that rely on amplitude cross-correlation similar to ones used in~\cite{Zhang:2022} for characterizing relative geometric accuracy between repeat passes. Figure~\ref{fig:offsets} shows the relative along-track offsets for two different stacks of WV mode data. We observe discrepancies up to 5 milliseconds (approximately 35 meters) in the along track direction. The range offsets are comparable to those observed in IW mode and have not been shown.

The observed along track shifts possibly indicate uncompensated timing corrections in the WV mode processor. We also note that WV mode was primarily designed to operate over open oceans, where relative geometric accuracy requirements are low and interferometry was not the primary target application for this mode.

\subsection{Stack generation}
\label{sec:coreg}
From Section~\ref{sec:geomcalib}, it is evident that geocoding each vignette independently, like we do in our IW mode burst-based pipelines~\cite{Agram:2022},  will not result in sub-pixel level coregistration necessary for interferometry. Hence, we adopted a modified workflow to generate coregistered stacks from WV mode vignettes. The modified approach is as follows:

\begin{enumerate}
    \item Geocode each vignette independently using its metadata and orbit information to an intermediate product at a posting of 2.5 meters.
    \item Estimate a bulk shift per image with respect to a reference image in the stack in map coordinates from the geocoded data using amplitude cross correlation and a large window of size greater than 1024, using a network-based approach~\cite{Fattahi:2017}. The estimated offsets in map coordinates are transformed to along-track time and slant range offset using a simple rotation matrix computed at the center of the correlation windows.
    \item Apply the estimated bulk offsets in along-track time and slant range to the vignette metadata and geocode them again to produce the final coregistered stack at a posting of 2.5 meters.
\end{enumerate}

The modified workflow implements the same two-stage coregistration paradigm, with coarse and fine coregistration steps, as traditional interferometric workflows~\cite{Sansosti:2006}, except that we perform these steps in map coordinates. Working directly in map coordinates allows us to eliminate the need to store and share large intermediate radar geometry grid-related arrays while benefiting from the simplified data management features within our platform~\cite{Beneke:2017}. The described workflow can be easily adopted to work with data from other SAR sensors and we have validated this approach by applying it to limited stacks of SLC data available to us from TerraSAR-X, COSMO SkyMed and Iceye as reported in~\cite{Agram:2022}.

Once WV mode data has been coregistered, geocoded and available on our platform, we are able to deploy our InSAR analytics pipelines that we have developed with IW mode~\cite{Olsen:2023} on this data. 

\begin{figure}[!h]
  \centering
  \includegraphics[height=10cm]{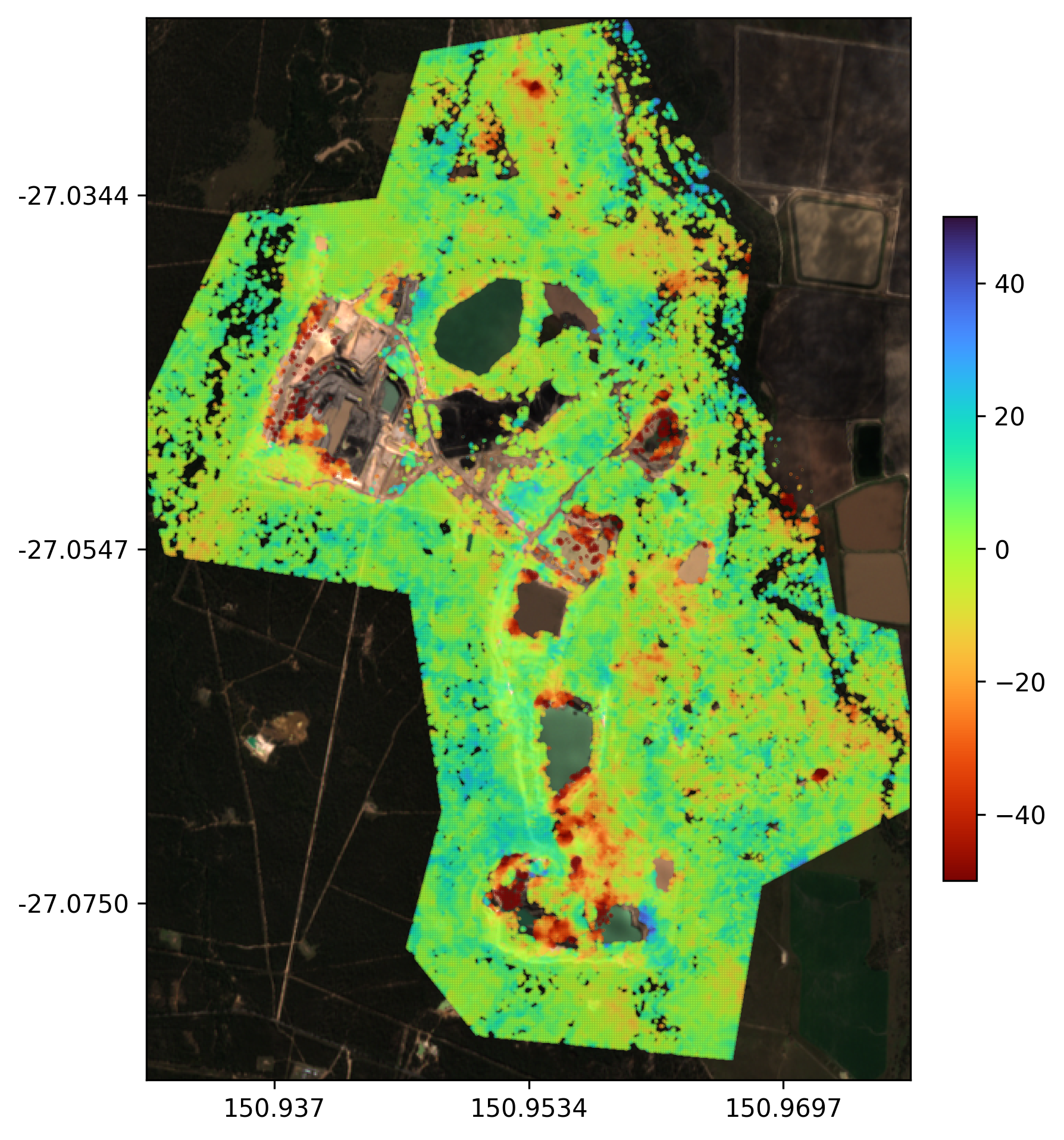}
  \caption{Scatter plot of points included in an InSAR deformation analysis based on complex WV mode data. Points are colored by cumulative deformation and overlaid on Sentinel-2 optical imagery. Deformation measurements are in millimeters.}
  \label{fig:site}
\end{figure}

\subsection{Results}
\label{sec:results}

\begin{figure}[!h]
  \centering
  \includegraphics[height=8cm]{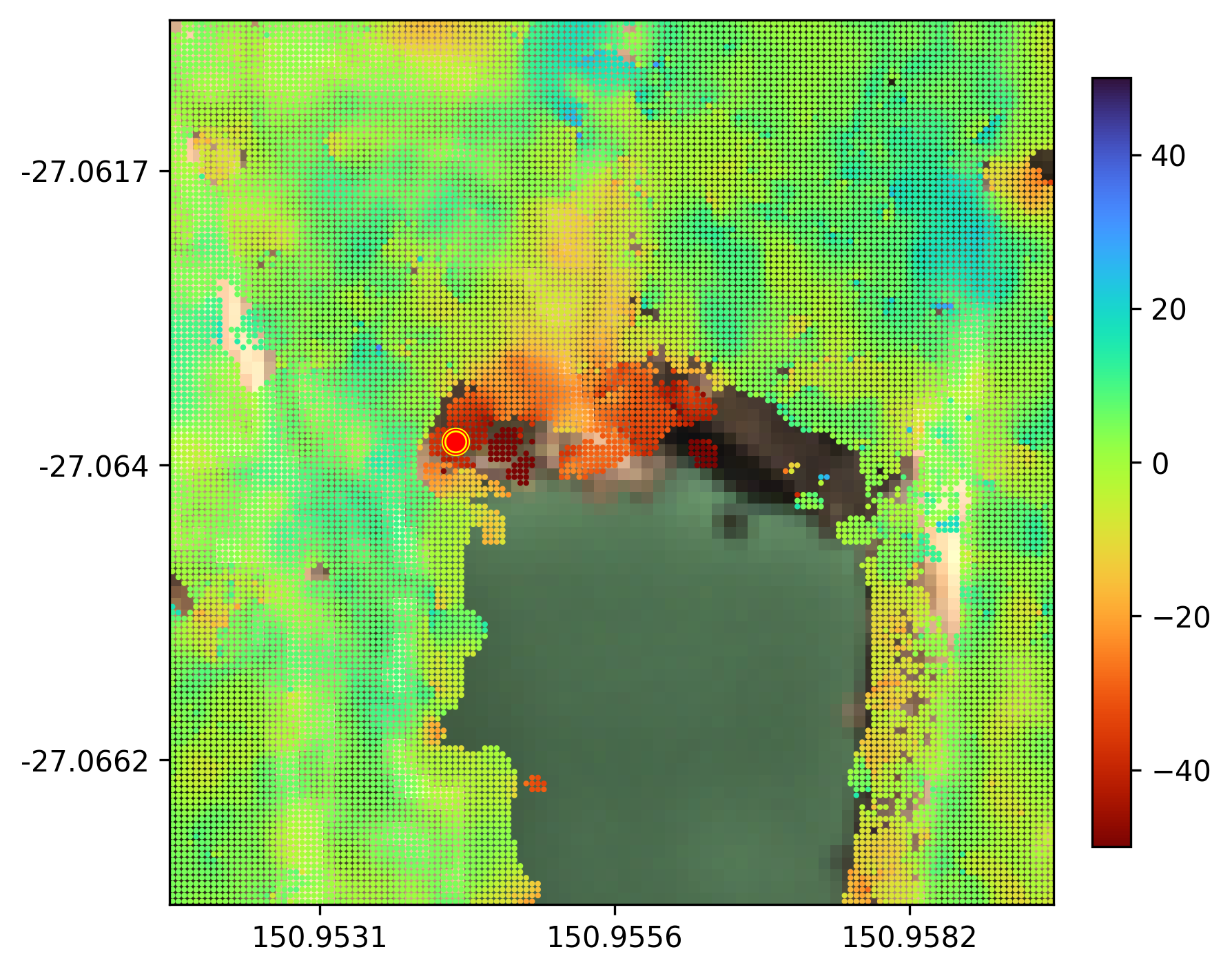}
  \caption{Cropped view of Fig~\ref{fig:site}. The thick red dot indicates the location where the deformation history is reported in Fig.~\ref{fig:deformation}.}
  \label{fig:detail}
\end{figure}

\begin{figure}[!h]
  \centering
  \includegraphics[height=8cm]{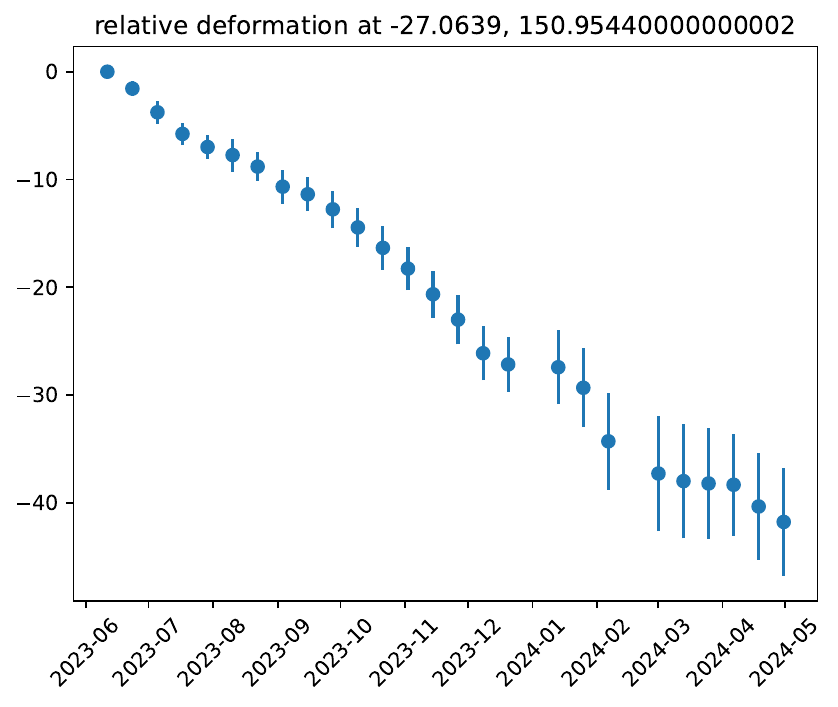}
  \caption{Deformation history in millimeters at the red dot in Fig~\ref{fig:detail}. Olsen et al. (\cite{Olsen:2023}) describes the method that generated the error bars.}
  \label{fig:deformation}
\end{figure}

In this section, we present the estimated deformation time-series estimated from a stack of Sentinel-1 WV mode data acquired over a mine in Dalby, Queensland, Australia from Track 38 from June 2023 to May 2024.  We processed the stack as described in Section~\ref{sec:coreg}, and we generated deformation estimates using the second method (Method 2) described in Section 4 of~\cite{Olsen:2023}. We analyzed the geocoded SLC  data on a downsampled 5 meter-by-5 meter grid as the high density of reliable scatterers still yielded over half a million points over this AOI. The results are shown in Figures~\ref{fig:site} and~\ref{fig:detail}. Figure~\ref{fig:deformation} shows the estimated relative deformation of a single pixel (the red dot in Figure~\ref{fig:deformation}).

We make the following observations from comparing results derived using WV mode data and IW mode data. 
\begin{itemize}
    \item With WV mode data we can generate deformation results on a 5m by 5m regular grid, while while IW mode data we generate results on a 10m by 10m regular grid. This factor of four allows us to resolve smaller deformation features than would be possible to resolve with IW data as input.
    \item The denser grid also results in increased resource usage and processing costs for the same physical area.
    \item The fraction of points selected when using the WV mode data as input was \emph{higher} than the fraction of points selected when using the IW mode data as input. We used the same coherence cutoff in both cases and we used roughly the same time range -- early June through late April 2024. This observation is consistent with previous studies~\cite{Ferretti:2011,Costantini:2017}. 
    \item It is worth reiterating the the SNR is lower for WV mode data, nevertheless, using these data as an input for an InSAR pipeline lead to results that were quite good compared to the results with the higher SNR IW mode data.
\end{itemize}

Overall, we are able to use Sentinel-1 WV mode data in our InSAR analytics pipelines in the same we would data acquired by other SAR sensors.

\section{Conclusion}
\label{sec:conclusion}

In this manuscript, we have described the salient features of WV mode SAR imagery acquired by the Sentinel-1 mission. We have described our approach of indexing individual WV mode vignettes, that allows us to work with this type of SAR data at scale. We have described the potential of WV mode backscatter-based applications over land and ocean.  We have also demonstrated the interferometric capabilities of Sentinel-1's WV mode with an example. We also demonstrated that Sentinel-1 WV mode data are potentially useful sources of data during emergency response and as a proxy for commercial smallsat SAR data for model and application development. With the imminent launch of Sentinel-1 C and D satellites, we expect the volume of WV mode acquisitions to increase again, including over land. While global coverage is not possible by design, and availability of deep stacks is not guaranteed with WV mode,  we will continue to treat this collection of SAR data similarly to other tasked  and opportunistic data collects from commercial providers for land and ocean applications.

\section*{Data Availability}
The single-look complex (SLC) Sentinel-1 imagery and associated metadata are available at the Alaska Satellite Facility’s Vertex Portal here: \url{https://search.asf.alaska.edu/#/} (accessed on 1 June 2024).

\section*{Acknowledgments}
The work presented in this manuscript was first conceptualized by Mike Warren, who also built the preliminary database of wave mode vignette footprints in 2020. This work contains Copernicus Sentinel-1 data (2016-2024) processed by European Space Agency.

\bibliographystyle{unsrt}  
\bibliography{references}

\begin{thebibliography}{10}

\bibitem{Torres:2012}
Ramon Torres, Paul Snoeij, Dirk Geudtner, David Bibby, Malcolm Davidson, Evert
  Attema, Pierre Potin, BjÖrn Rommen, Nicolas Floury, Mike Brown,
  Ignacio~Navas Traver, Patrick Deghaye, Berthyl Duesmann, Betlem Rosich, Nuno
  Miranda, Claudio Bruno, Michelangelo L'Abbate, Renato Croci, Andrea
  Pietropaolo, Markus Huchler, and Friedhelm Rostan.
\newblock Gmes sentinel-1 mission.
\newblock {\em Remote Sensing of Environment}, 120:9 -- 24, 2012.
\newblock The Sentinel Missions - New Opportunities for Science.

\bibitem{Ferretti:2021}
Alessandro Ferretti, Emanuele Passera, and Renalt Capes.
\newblock End-to-end implementation and operation of the {European Ground
  Motion Service (EGMS): Algorithm Theoretical Basis Document}.
\newblock Technical Report EGMS-D3-ALG-SC1-2.0-006, European Environment
  Agency, dec 2021.

\bibitem{Agram:2022}
Piyush~S. Agram, Michael~S. Warren, Matthew~T. Calef, and Scott~A. Arko.
\newblock An efficient global scale {Sentinel-1} radar backscatter and
  interferometric processing system.
\newblock {\em Remote Sensing}, 14(15), 2022.

\bibitem{Lehner:2000}
S.~Lehner, J.~Schulz-Stellenfleth, B.~Schattler, H.~Breit, and J.~Horstmann.
\newblock Wind and wave measurements using complex {ERS-2 SAR} wave mode data.
\newblock {\em IEEE Transactions on Geoscience and Remote Sensing},
  38(5):2246--2257, 2000.

\bibitem{Hasselmann:2013}
Klaus Hasselmann, B~Chapron, L~Aouf, F~Ardhuin, F~Collard, G~Engen, Susanne
  Hasselmann, P~Heimbach, H~Johnsen, et~al.
\newblock The {ERS SAR} wave mode: A breakthrough in global ocean wave
  observations.
\newblock 2013.

\bibitem{S1userguide:2023}
{{European Space Agency}}.
\newblock {{Sentinel-1 User SAR Guide}}.
\newblock
  \url{https://sentinel.esa.int/web/sentinel/user-guides/sentinel-1-sar}, 2023.
\newblock Accessed on Jan 1, 2024.

\bibitem{Hajduch:2022}
Guillaume Hajduch, Matthieu Bourbigot, Harald Johnsen, and Riccardo Piantanida.
\newblock Sentinel-1 level 1 detailed algorithm definition.
\newblock Technical Report S1-RS-MDA-52-7441, European Space Agency, feb 2022.

\bibitem{Burst:2022}
{{Sentinel-1 Mission Performance Cluster}}.
\newblock Sentinel-1 burst id map, version 20220530.
\newblock \url{https://sar-mpc.eu/test-data-sets/}, may 2022.
\newblock Generated by Sentinel-1 SAR MPC.

\bibitem{Ardhuin:2015}
Fabrice Ardhuin, Fabrice Collard, Bertrand Chapron, Fanny Girard-Ardhuin,
  Gilles Guitton, Alexis Mouche, and Justin~E. Stopa.
\newblock Estimates of ocean wave heights and attenuation in sea ice using the
  sar wave mode on sentinel-1a.
\newblock {\em Geophysical Research Letters}, 42(7):2317--2325, 2015.

\bibitem{Stopa:2017}
J.~E. Stopa and A.~Mouche.
\newblock Significant wave heights from sentinel-1 sar: Validation and
  applications.
\newblock {\em Journal of Geophysical Research: Oceans}, 122(3):1827--1848,
  2017.

\bibitem{Li:2019}
Huimin Li, Alexis Mouche, Justin~E. Stopa, and Bertrand Chapron.
\newblock Calibration of the normalized radar cross section for {Sentinel-1}
  wave mode.
\newblock {\em IEEE Transactions on Geoscience and Remote Sensing},
  57(3):1514--1522, 2019.

\bibitem{Wang:2019}
Chen Wang, Alexis Mouche, Pierre Tandeo, Justin~E Stopa, Nicolas
  Long{\'e}p{\'e}, Guillaume Erhard, Ralph~C Foster, Douglas Vandemark, and
  Bertrand Chapron.
\newblock A labelled ocean {SAR} imagery dataset of ten geophysical phenomena
  from {Sentinel-1} wave mode.
\newblock {\em Geoscience Data Journal}, 6(2):105--115, 2019.

\bibitem{Mirandageo:2017}
Nuno Miranda, Peter Meadows, Riccardo Piantanida, Andrea Recchia, David Small,
  Adrian Schubert, Pauline Vincent, Dirk Geudtner, Ignacio Navas-Traver, and
  Francisco~Ceba Vega.
\newblock The sentinel-1 constellation mission performance.
\newblock In {\em 2017 IEEE International Geoscience and Remote Sensing
  Symposium (IGARSS)}, pages 5541--5544, 2017.

\bibitem{Zhang:2022}
Zhang Yunjun, Heresh Fattahi, Xiaoqing Pi, Paul Rosen, Mark Simons, Piyush
  Agram, and Yosuke Aoki.
\newblock Range geolocation accuracy of c-/l-band sar and its implications for
  operational stack coregistration.
\newblock {\em IEEE Transactions on Geoscience and Remote Sensing}, 60:1--19,
  2022.

\bibitem{Fattahi:2017}
Heresh Fattahi, Piyush Agram, and Mark Simons.
\newblock A network-based enhanced spectral diversity approach for tops
  time-series analysis.
\newblock {\em IEEE Transactions on Geoscience and Remote Sensing},
  55(2):777--786, 2017.

\bibitem{Sansosti:2006}
Eugenio Sansosti, Paolo Berardino, Michele Manunta, Francesco Serafino, and
  Gianfranco Fornaro.
\newblock Geometrical sar image registration.
\newblock {\em IEEE Transactions on Geoscience and Remote Sensing},
  44(10):2861--2870, 2006.

\bibitem{Beneke:2017}
Carly~Marie Beneke, Samuel Skillman, Michael~S Warren, Tim Kelton,
  Steven~Patrick Brumby, Rick Chartrand, and Mark Mathis.
\newblock A platform for scalable satellite and geospatial data analysis.
\newblock In {\em AGU Fall Meeting Abstracts}, volume 2017, pages IN32C--04,
  2017.

\bibitem{Olsen:2023}
Kelly~M Olsen, Matthew~T Calef, and Piyush~S Agram.
\newblock Contextual uncertainty assessments for insar-based deformation
  retrieval using an ensemble approach.
\newblock {\em Remote Sensing of Environment}, 287:113456, 2023.

\bibitem{Ferretti:2011}
Alessandro Ferretti, Andrea Tamburini, Fabrizio Novali, Alfio Fumagalli,
  Giacomo Falorni, and Alessio Rucci.
\newblock Impact of high resolution radar imagery on reservoir monitoring.
\newblock {\em Energy Procedia}, 4:3465--3471, 2011.
\newblock 10th International Conference on Greenhouse Gas Control Technologies.

\bibitem{Costantini:2017}
Mario Costantini, Alessandro Ferretti, Federico Minati, Salvatore Falco,
  Francesco Trillo, Davide Colombo, Fabrizio Novali, Fabio Malvarosa, Claudio
  Mammone, Francesco Vecchioli, Alessio Rucci, Alfio Fumagalli, Jacopo Allievi,
  Maria~Grazia Ciminelli, and Salvatore Costabile.
\newblock Analysis of surface deformations over the whole italian territory by
  interferometric processing of ers, envisat and cosmo-skymed radar data.
\newblock {\em Remote Sensing of Environment}, 202:250--275, 2017.
\newblock Big Remotely Sensed Data: tools, applications and experiences.

\end{thebibliography}

\end{document}